\documentclass[
superscriptaddress,aps,preprintnumbers,amsmath,showpacs,amssymb,prl,reprint,nofootinbib]{revtex4-1}

\usepackage[dvipdfm]{graphicx}
\usepackage{amsfonts}
\usepackage[mathscr]{eucal}
\usepackage{ascmac}
\usepackage{epstopdf}
\usepackage{dcolumn}
\usepackage{bm}
\usepackage{hyperref}
\usepackage{color}

\begin{document}


\def\a{\alpha}
\def\b{\beta}
\def\c{\varepsilon}
\def\d{\delta}
\def\e{\epsilon}
\def\f{\phi}
\def\g{\gamma}
\def\h{\theta}
\def\k{\kappa}
\def\l{\lambda}
\def\m{\mu}
\def\n{\nu}
\def\p{\psi}
\def\q{\partial}
\def\r{\rho}
\def\s{\sigma}
\def\t{\tau}
\def\u{\upsilon}
\def\v{\varphi}
\def\w{\omega}
\def\x{\xi}
\def\y{\eta}
\def\z{\zeta}
\def\D{\Delta}
\def\G{\Gamma}
\def\H{\Theta}
\def\L{\Lambda}
\def\F{\Phi}
\def\P{\Psi}
\def\S{\Sigma}

\def\o{\over}
\def\beq{\begin{eqnarray}}
\def\eeq{\end{eqnarray}}
\newcommand{\gsim}{ \mathop{}_{\textstyle \sim}^{\textstyle >} }
\newcommand{\lsim}{ \mathop{}_{\textstyle \sim}^{\textstyle <} }
\newcommand{\vev}[1]{ \left\langle {#1} \right\rangle }
\newcommand{\bra}[1]{ \langle {#1} | }
\newcommand{\ket}[1]{ | {#1} \rangle }
\newcommand{\EV}{ \ {\rm eV} }
\newcommand{\KEV}{ \ {\rm keV} }
\newcommand{\MEV}{\  {\rm MeV} }
\newcommand{\GEV}{\  {\rm GeV} }
\newcommand{\TEV}{\  {\rm TeV} }
\newcommand{\1}{\mbox{1}\hspace{-0.25em}\mbox{l}}
\def\diag{\mathop{\rm diag}\nolimits}
\def\Spin{\mathop{\rm Spin}}
\def\SO{\mathop{\rm SO}}
\def\O{\mathop{\rm O}}
\def\SU{\mathop{\rm SU}}
\def\U{\mathop{\rm U}}
\def\Sp{\mathop{\rm Sp}}
\def\SL{\mathop{\rm SL}}
\def\tr{\mathop{\rm tr}}

\def\IJMP{Int.~J.~Mod.~Phys. }
\def\MPL{Mod.~Phys.~Lett. }
\def\NP{Nucl.~Phys. }
\def\PL{Phys.~Lett. }
\def\PR{Phys.~Rev. }
\def\PRL{Phys.~Rev.~Lett. }
\def\PTP{Prog.~Theor.~Phys. }
\def\ZP{Z.~Phys. }

\def\dd{\mathrm{d}}
\def\ff{\mathrm{f}}
\def\BH{{\rm BH}}
\def\inf{{\rm inf}}
\def\ev{{\rm evap}}
\def\eq{{\rm eq}}
\def\SM{{\rm sm}}
\def\Mpl{M_{\rm Pl}}
\def\GeV{ \ {\rm GeV}}
\newcommand{\Red}[1]{\textcolor{red}{#1}}

\def\mDM{m_{\rm DM}}
\def\TeV{\ {\rm TeV}}
\def\MeV{\ {\rm MeV}}
\def\Gphi{\Gamma_\phi}
\def\TR{T_{\rm RH}}
\def\Br{{\rm Br}}
\def\DM{{\rm DM}}
\def\Eth{E_{\rm th}}
\newcommand{\lmk}{\left(}  
\newcommand{\rmk}{\right)}
\newcommand{\lkk}{\left[}  
\newcommand{\rkk}{\right]}
\newcommand{\lhk}{\left \{ }  
\newcommand{\rhk}{\right \} }
\newcommand{\del}{\partial}  
\newcommand{\la}{\left\langle} 
\newcommand{\ra}{\right\rangle}

\newcommand{\qel}{\hat{q}_{el}}
\newcommand{\ksplit}{k_{\text{split}}}
\def\GDM{\Gamma_{\text{DM}}}
\newcommand{\half}{\frac{1}{2}}
\def\Gsplit{\Gamma_{\text{split}}}

\def\mg{m_{3/2}}
\newcommand{\abs}[1]{\left\vert {#1} \right\vert}
\def\Im{{\rm Im}}
\def\bea{\begin{array}}
\def\eea{\end{array}}
\def\Mpl{M_{\text{Pl}}}
\def\M{M_{\text{Pl}}}
\def\mN{m_{\text{NLSP}}}
\def\Td{T_{\text{decay}}}
\def\mphi{m_{\phi}}
\def\tanb{\text{tan}\beta}
\def\signmu{\text{sign}[\mu]}
\def\fb{\text{ fb}}
\def\ij{_{ij}}
\def\k{\lmk {\bf k} \rmk}
\def\tk{\lmk \tau, {\bf k} \rmk}
\def\xk{\lmk x, {\bf k} \rmk}
\def\TT{T_{ij}^{\rm TT}}
\def\Hz{\ {\rm Hz}}
\def\for{\quad \text{for }}
\def\Min{\text{Min}}
\def\Max{\text{Max}}


\title{
Gravitational waves as a probe of SUSY scale 
}

\author{Ayuki Kamada}
\affiliation{Kavli IPMU (WPI), TODIAS, University of Tokyo, Kashiwa, 277-8583, Japan}
\affiliation{Department of Physics and Astronomy, University of California, Riverside, CA, 92507, USA}
\author{Masaki Yamada}
\affiliation{Kavli IPMU (WPI), TODIAS, University of Tokyo, Kashiwa, 277-8583, Japan}
\affiliation{ICRR, University of Tokyo, Kashiwa, 277-8582, Japan}
\begin{abstract}
We investigate the sources of the Hubble-induced mass 
for a flat direction in supersymmetric theories 
and show that 
the sign of the Hubble-induced mass generally changes just after the end of inflation. 
This implies that 
global cosmic strings generally form after the end of inflation in most supersymmetric models, 
including the Minimal Supersymmetric Standard Model. 
The cosmic strings emit gravitational waves 
whose frequency corresponds to the Hubble scale, 
until they disappear when the Hubble parameter decreases down to the soft mass of the flat direction. 
As a result, 
the peak frequency of gravitational waves is related to the supersymmetric scale. 
The observation of this gravitational wave signal will give us 
information of supersymmetric scale and reheating temperature. 
\end{abstract}

\date{\today}
\maketitle
\preprint{IPMU 14-0158}
\preprint{ICRR-Report-685-2014-11}

\noindent {\bf Introduction. }
The observation of gravitational waves (GWs) will open a new window onto the early Universe 
and provide information on physics at correspondingly high energy scales~\cite{Maggiore:1999vm}. 
Stochastic GW signals are generated by non-equilibrium phenomena during the post-inflationary period, 
such as preheating%
~\cite{
Khlebnikov:1997di, Easther:2006gt, Easther:2006vd, GarciaBellido:2007dg, 
GarciaBellido:2007af, Dufaux:2007pt, Dufaux:2008dn%
}, 
first order phase transition%
~\cite{
Witten:1984rs, Kosowsky:1992rz, Apreda:2001us, Nicolis:2003tg, 
Grojean:2006bp, Caprini:2007xq, Caprini:2009fx, Huber:2008hg, Hindmarsh:2013xza%
}, 
turbulent motions%
~\cite{
Kosowsky:2001xp, Dolgov:2002ra, Gogoberidze:2007an, Caprini:2006jb%
}, 
topological defects%
~\cite{
Vilenkin:1981bx, Caldwell:1991jj, Vachaspati:1984gt, Olmez:2010bi, Dufaux:2010cf, 
Kawasaki:2011vv, Figueroa:2012kw, Hiramatsu:2013qaa%
}, 
and self-ordering scalar fields%
~\cite{
Turok:1991qq, Krauss:1991qu, JonesSmith:2007ne, Fenu:2009qf, Durrer:2014raa%
}. 
Quantum fluctuations during inflation are another source of 
GWs, called the inflationary GW background~\cite{Starobinsky:1979ty, Rubakov:1982df}. 
Each source may predict characteristic GW signals 
measureable by future GW detectors; 
such as LISA~\cite{lisa}, DECIGO~\cite{decigo}, Advanced LIGO~\cite{adv_ligo}, and ET~\cite{et}. 
The observation of these GW signals 
will improve our understanding on particle physics beyond the Standard Model.

Supersymmetric (SUSY) theories are well-motivated in particle physics, 
because it addresses the hierarchy problem 
and also achieves gauge coupling unification. 
In SUSY theories, 
there usually exist scalar fields 
called flat directions, 
whose potentials are 
flat as long as they maintain SUSY and renormalizability (see Ref.~\cite{Gherghetta:1995dv}). 
In this letter, 
we investigate the dynamics of a flat direction 
and show that cosmic string network gerenally forms 
after the end of inflation. 
They eventually disappear when the Hubble parameter decreases 
down to the mass of the flat direction. 
Since 
the cosmic strings generate a stochastic GW background 
with a peak frequency corresponding to the Hubble scale, 
the information of the mass of the flat direction is imprinted on 
the GW spectrum. 
This mechanism will give us information of SUSY scale 
through GW detection 
even if its energy scale is beyond the reach of 
future accelerators.

\vspace{0.2cm}
\noindent {\bf Dynamics of flat direction. }
Let us focus on one flat direction, which we denote as $\phi$. 
During inflation, 
the flat direction obtains Hubble-induced mass 
through higher-dimensional K\"{a}hler potentials like 
\beq
 \int \dd^2 \theta \dd^2 \bar{\theta} \frac{c'_{H_{\rm inf}}}{M_*^2} \abs{X}^2 \abs{\phi}^2, 
\eeq
where $M_*$ is a cut-off scale, and $c'_{H_{\rm inf}}$ is an $O(1)$ constant. 
We expect that 
$M_*$ is less than or equal to the Planck scale $\M$ ($\simeq 2.4 \times 10^{18} \GeV$). 
The $F$-term of $X$ drives inflation 
and satisfies the relation of $\abs{F_X}^2 = 3 H_{\rm inf}^2 \M^2$. 
This implies that 
the flat direction obtains a Hubble-induced mass of $c_{H_{\rm inf}} H_{\rm inf}^2 \abs{\phi}^2$ during inflation, 
where 
\beq
 c_{H_{\rm inf}} = 3 c'_{H_{\rm inf}} \frac{\M^2}{M_*^2}. 
\eeq
While the coefficient $c_{H_{\rm inf}}$ is assumed to be negative 
in the context of the Affleck-Dine baryogenesis~\cite{AD, DRT}, 
we assume $c_{H_{\rm inf}} > 0$ in this paper. 
In this case, the flat direction stays at the origin, i.e., $\phi = 0$, during inflation.

After inflation ends and before reheating completes, 
the energy density of the Universe is 
dominated by the oscillation of a scalar field (denoted by $I$) 
and the Hubble parameter decreases with time as $H (t) \propto a^{-3/2} (t)$, 
where $a(t)$ is the scale factor. 
During this oscillation era, the flat direction obtains a Hubble-induced mass
through 
\beq
  \int \dd^2 \theta \dd^2 {\bar \theta} 
 \frac{c'_H}{M_*^2} \abs{I}^2 \abs{\phi}^2 
\supset
 \frac{c'_H}{M_*^2}  \abs{\dot{I}}^2 \abs{\phi}^2. 
\eeq
Since the oscillation time scale of $I$ is much smaller than $H^{-1}$, 
we can average $|\dot{I}|^2$ over the oscillation time scale like 
$| \dot{I}|^2  \approx \rho_I (t)/2 \simeq (3/2)  H^2 (t) \M^2$. 
Thus, we obtain the Hubble induced mass of $c_H H^2 (t) \abs{\phi}^2$ with 
\beq
 c_{H} = \frac{3}{2} c'_{H} \frac{\M^2}{M_*^2}, 
 \label{c_H}
\eeq
during the oscillation era. 
The coefficient of Hubble-induced mass during inflation, 
$c_{H_{\rm inf}}$, is generally different from the one during the oscillation era, $c_H$, 
because the field $I$ is generally different from the field $X$. 
For example, 
in the chaotic inflation model proposed in Ref.~\cite{Kawasaki:2000yn}, 
the field $X$ (in this letter) is identified with the field $X$ (in Ref.~\cite{Kawasaki:2000yn})
while $I$ is identified with the inflaton $\varphi$. 
In the simplest hybrid inflation model 
proposed in Ref.~\cite{Copeland:1994vg} (Ref.~\cite{Dvali:1994ms}), 
the field $X$ is identified with the field $\Phi$ ($S$) 
while $I$ is a water-fall field 
$\Psi_1$ and $\Psi_2$ ($\phi$ and $\bar{\phi}$). 
While the non-renormalizable terms are heavily dependent on 
the high-energy physics beyond the cut-off scale, 
we assume that among many flat directions in SUSY theories (at least) 
one flat direction has $c_{H_{\rm inf}} >0$ and $c_H < 0$. 
Let us investigate the dynamics of such a flat direction. 
The flat direction 
stays at $\phi=0$ during inflation, and then, 
it obtains a large VEV after the end of inflation.

If the flat direction has a non-renormalizable superpotential, 
the resultant GW background is generally too small to be detected~\cite{full paper}. 
Thus, we consider the case that 
the superpotential of the flat direction is absent due to discrete R-symmetries. 
In this case, 
the potential of the flat direction obtains higher-dimentional terms 
coming from non-renormalizable K\"{a}hler potential 
and is written as 
\beq
 V (\phi) = 
 c_H H^2(t) \abs{\phi}^2 
 + \mphi^2 \abs{\phi}^2 
 + a_{H} H^2(t) \frac{\abs{\phi}^{2n-2}}{\M^{2n-4}} 
 + \dots,
 \label{potential2}
\eeq
where $n$ ($\ge 3$) is a certain integer, $\mphi$ is the soft mass of the flat direction, 
and $a_{H}$ is given by 
\beq
 a_H = a_H' \lmk \frac{\M}{M_*} \rmk^{2n-2}, 
 \label{a_H}
\eeq
with $a_H'$ being an $O(1)$ constant. 
The dots $\dots$ represent the other irrelevant higher-dimentional terms. 
Here we assume that higher-dimentional terms breaking $U(1)$ symmetry like $\phi^n + \text{c.c.}$ 
are absent due to R-symmetries. 
The potential minimum is determined as 
\beq
 \la \phi \ra = 
 \lmk \frac{\abs{c_H}}{a_{H} (n-1)} \rmk^{1/(2n-4)} \M \ \  \lmk \sim M_* \rmk, 
 \label{VEV}
\eeq
as long as $c_H H^2(t) \gg \mphi^2$.

Here we investigate the components of the flat direction in detail. 
Let us consider $L_i H_u$ flat direction as an illustration, 
where $i$ is a flavour index.%
\footnote{
The absence of the non-renormalizable superpotential for $L_i H_u$ is 
disfavored from the viewpoint of the observed neutrino oscillations. 
In this letter, however, we take $L_i H_u$ flat direction as an example to take 
advantage of its simple flavour structure. 
}
Since the above non-renormalizable K\"{a}hler potentials generally breaks flavour symmetry, 
$L_i H_u$ flat direction has the local SU(2)$_L$$\times$U(1)$_Y$ symmetry 
and the global U(1)$_L$ symmetry. 
The non-zero VEV of the flat direction breaks 
the local SU(2)$_L$$\times$U(1)$_Y$ symmetry 
to the U(1)$_{\rm EM}$ symmetry 
and breaks the global U(1)$_L$ symmetry completely. 
As a result, global cosmic string network forms after the end of inflation.%
\footnote{
If the flat direction has a larger flavour symmetry like SU(3)$_{\rm flavour}$, 
the non-zero VEV of the flat direction breaks the global 
SU(3)$_{\rm flavour} \times$U(1)$_L$ symmetry to the SU(2)$\times$U(1) symmetry. 
Also in this case, 
GWs are emitted by the dynamics of randomly distributed Nambu-Goldstone modes 
of the flat direction~\cite{Turok:1991qq, Krauss:1991qu, 
JonesSmith:2007ne, Fenu:2009qf, Durrer:2014raa, full paper}, 
though cosmic strings are absent.%
}
By using numerical simulations (see below), 
we confirm that cosmic string network reaches a scaling regime within a certain time. 
In this regime, the number of cosmic strings in the Hubble volume is $O(1)$.

Since the Hubble-induced mass decreases with time as $\propto H (t) \propto a^{-3/2} (t)$, 
the soft mass $\mphi$ eventually dominates 
the potential of the flat direction 
and the flat direction starts to oscillate around $\phi = 0$. 
Let us denote that time as $t_{\rm decay}$, 
which is estimated by 
\beq
 c_H H^2 (t_{\rm decay}) \simeq \mphi^2. 
 \label{t_decay}
\eeq
Note that for low-scale SUSY models 
we should require $\TR \lesssim 10^9 \GeV$ 
to avoid the gravitino problem~\cite{Kawasaki:2004qu, Kawasaki:2008qe}. 
Even for high-scale SUSY models such as pure gravity mediation, 
$\TR \lesssim 10^{10} \GeV$ is required 
to avoid an overproduction of LSP ($= O(100) \GeV$)~\cite{Ibe:2004tg, Ibe:2004gh, Ibe:2011aa}. 
In such well-motivated cases, 
the cosmic strings dissapear before reheating completes, 
that is, $t_{\rm decay} < t_{\rm RH}$, where $t_{\rm RH}$ is the time reheating completes. 
Hereafter we consider such a case.

\vspace{0.2cm}
\noindent {\bf Caluculation of GWs. }
GWs are emitted by the dynamics of the cosmic strings%
~\cite{
Vilenkin:1981bx, Caldwell:1991jj, Vachaspati:1984gt, Olmez:2010bi, 
JonesSmith:2007ne, Fenu:2009qf, Figueroa:2012kw%
}. 
We calculate the spectrum of GWs using the method proposed in Ref.~\cite{Dufaux:2007pt, Kawasaki:2011vv}, 
which is suitable to our situation compared with the method to calculate GW amplitudes from localized sources 
derived in Ref.~\cite{Weinberg}.%
\footnote{
Since $\dd^2 V (\phi) / \dd \phi^2 \sim H^2 (t)$ at $\phi = \la \phi \ra$, 
a typical width of cosmic strings is of the order of the Hubble radius. 
This implies that 
the Nambu-Goto approximation, which was used in Ref.~\cite{Albrecht:1984xv} for example, 
is inappropriate to describe these cosmic strings. 
}
Hereafter, we change the time variable from $t$ to the conformal time $\tau$ 
which is defined by $\dd t = a \dd \tau$.

The energy density of GWs can be calculated from~\cite{Dufaux:2007pt} 
\beq
\Omega_{\rm gw} (\tau) 
 &\equiv& 
 \frac{1}{\rho_{\rm tot} \lmk \tau \rmk} \frac{ \dd \rho_{\rm gw} ( \tau) }{\dd \log k} \nonumber\\ 
 &\simeq& 
 \frac{k^5}{24 V a^4 H^2} 
 \int \dd \Omega_k
 \sum\ij \lmk \abs{A\ij }^2 + \abs{B\ij }^2 \rmk, 
 \label{omega_gw}
\eeq
where $\rho_{\rm tot} (\tau )$ ($= 3 \M^2 H^2 (\tau)$) is the total energy density of the Universe. 
The frequency-dependent coefficients $A_{ij}$ and $B\ij$ are given as~\cite{Kawasaki:2011vv} 
\beq
 A\ij \k &=& - 16 \pi G \int_{\tau_i}^{\tau_f} \dd \tau' \  \tau' a(\tau') f_A (k \tau') \TT \lmk \tau', {\bf k} \rmk, \label{A}\\
 B\ij \k &=& 16 \pi G \int_{\tau_i}^{\tau_f} \dd \tau' \  \tau' a(\tau') f_B (k \tau' ) \TT \lmk \tau', {\bf k} \rmk, \label{B}
\eeq
where 
$\TT$ is the Fourier transformed transverse-traceless part of the anisotropic stress. 
Here we have implicitly assumed that the source term $T^{TT}_{ij}$ 
lasts during the interval of $[\tau_i, \tau_f]$. 
The functions $f_A$ and $f_B$ are 
calculated by matching solutions of the Einstein equation in the oscillation era 
with that in the radiation dominated era at $\tau = \tau_{\rm RH}$
such as~\cite{full paper}
\beq
f_A (k \tau') &=& 
 \lkk a_1 n_1(k \tau') - a_2 j_1(k \tau') \rkk, \nonumber\\
f_B (k \tau') &=& 
 \lkk -b_1 n_1(k \tau') + b_2 j_1(k \tau') \rkk, 
 \label{functions}
\eeq
where 
$j_l$ and $n_l$ are the spherical Bessel and Neumann functions of order $l$, 
and the coefficients are given as 
\beq
 a_1 &=& \left. x^2 \lkk j_1 (x) \del_x n_0 (x) - n_0 (x) \del_x j_1 (x)  \rkk \right. \vert_{x \to k \tau_{\rm RH}}, 
 \nonumber\\ 
 a_2 &=& \left. x^2 \lkk n_1 (x) \del_x n_0 (x) - n_0 (x) \del_x n_1 (x)  \rkk \right. \vert_{x \to k \tau_{\rm RH}}, 
\nonumber\\
 b_1 &=& - \left. x^2 \lkk j_1 (x) \del_x j_0 (x) - j_0 (x) \del_x j_1 (x)  \rkk \right. \vert_{x \to k \tau_{\rm RH}}, 
\nonumber\\
 b_2 &=& - \left. x^2 \lkk n_1 (x) \del_x j_0 (x) - j_0 (x) \del_x n_1 (x)  \rkk \right. \vert_{x \to k \tau_{\rm RH}}. 
\label{coefficients}
\eeq

\begin{figure}[t]
\centering 
\includegraphics[width=.45\textwidth, bb=0 0 360 347]{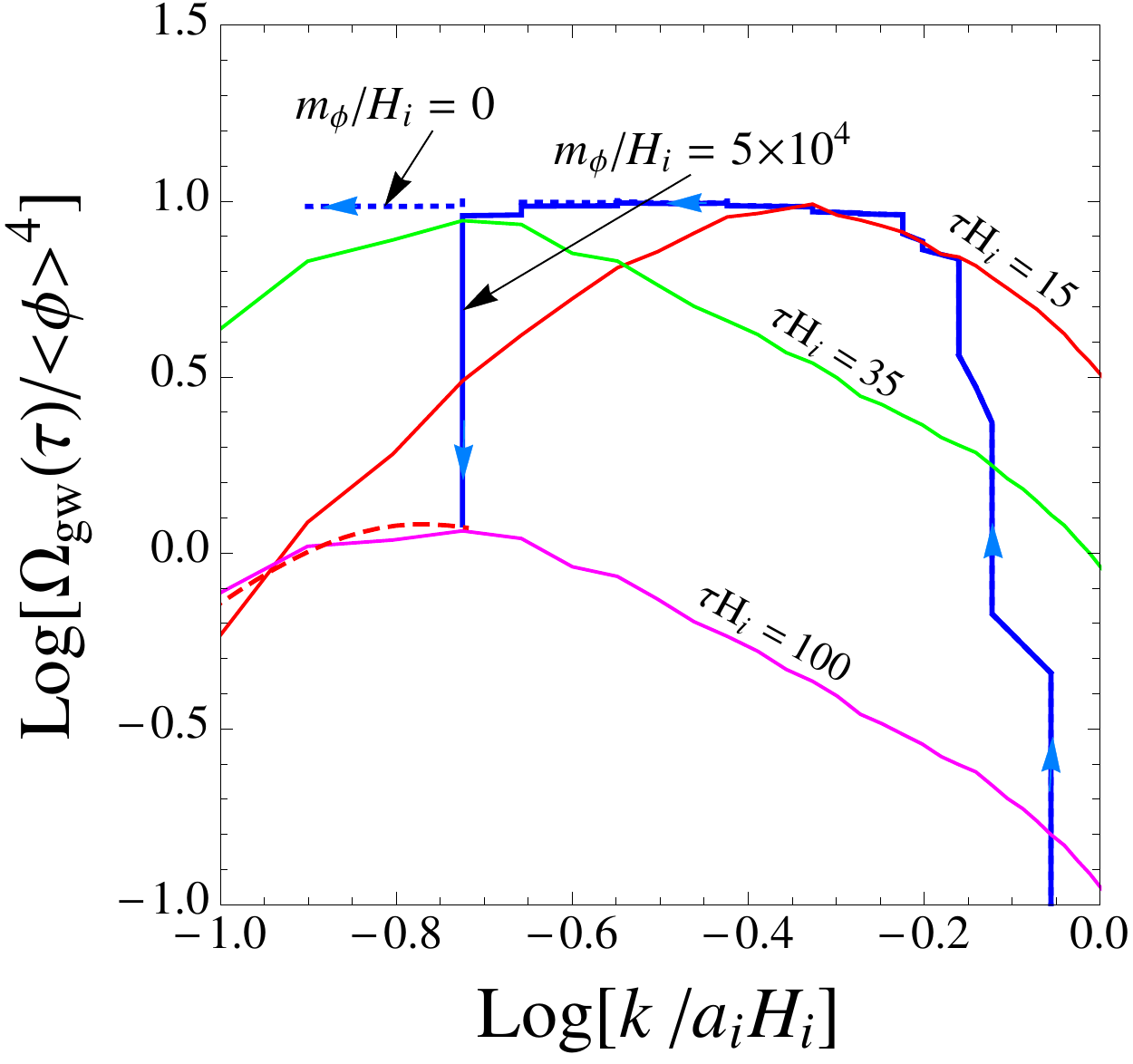} 
\caption{
Evolution of GW spectra obtained by numerical calculations. 
We show the obtained spectra 
at $\tau H_i=15$ (red line), $\tau H_i= 35$ (green line), and $\tau H_i= 100$ (magenta line). 
We take $n=4$, $c_H =15$, and $\mphi / H_i= 5 \times 10^{-4}$. 
The blue (dotted) curve represents the contour of the peak frequency and the peak GW energy density
for the case of $\mphi / H_i= 5 \times10^{-4}$ ($0$). 
The red dashed curve represents an analitic estimation given by Eqs.~(\ref{A}) and (\ref{B}) 
with $k$-independent $\TT$. 
}
  \label{fig1}
\end{figure}

Here let us estimate the spectrum of GWs from the cosmic strings
before we show our simulation results. 
GWs are most efficiently emitted for the frequency corresponding to the Hubble radius, 
which is given as $k \sim \tau^{-1}$. 
Since the emission of GWs proceeds through the Planck suppressed interaction 
(see Eqs.~(\ref{omega_gw}), (\ref{A}), and (\ref{B})), 
the produced energy density of GWs can be estimated as%
\footnote{
If the cut-off scale is equal to the Planck scale ($M_* \simeq \M$), 
the right hand side of Eq.~(\ref{omega_gw3}) becomes $O(1)$, 
which implies that we have to include the effect of the backreaction of GW emission. 
In addition, the energy density of the cosmic strings is comparable to 
the energy density of the Universe and the evolution of the Universe is nontrivial. 
Hereafter, 
we consider the case that the cut-off scale is less than the Planck scale. 
}
\beq
 \frac{\Delta \Omega_{\rm gw} }{\Delta \log \tau} 
 \sim 
 \lmk \frac{\la \phi \ra}{\M} \rmk^4 
 \sim 
 \lmk \frac{M_*}{\M} \rmk^4, 
 \label{omega_gw3}
\eeq
where we use Eq.~(\ref{VEV}). 
The GW energy density decreases with time as $\propto a^{-1} (\tau) \propto \tau^{-2}$ in the oscillation era 
and the peak frequency of the emitted GWs decreases with time as $\propto \tau^{-1}$. 
Therefore, the GW spectrum for the frequency $k \gtrsim \tau^{-1}$ is proportional to $k^{-2}$.%
\footnote{
Although 
our simulation results (Fig.~\ref{fig1}) appears different from $k^{-2}$ for $k \gtrsim \tau^{-1}$, 
we have confirmed that this is owing to the limitation of the simulation time (see Ref.~\cite{full paper}). 
}
For large-scale modes $k \lesssim \tau^{-1}$, 
the Fourier transformed transverse-traceless part of the anisotropic stress $\TT$ 
is independent of $k$ due to the loss of causality at the large scale~\cite{Dufaux:2007pt, Kawasaki:2011vv}. 
Then, using $j_l(x) \to  2^l l!  x^l/(2l+1)!$ and $n_l(x) \to -(2l)!/ (2^l l! x^{l+1})$ for $x \ll 1$, 
we obtain $\Omega_{\rm gw} \propto k$ for $\tau_{\rm RH}^{-1} \ll k \ll \tau^{-1}$ 
and $\Omega_{\rm gw} \propto k^3$ for $k \ll \tau_{\rm RH}^{-1}$. 
Thus, the spectrum of GWs bends at the wavenumber around $k \simeq \tau_{\rm RH}^{-1}$.

The GW emission terminates at the time of $\tau_{\rm decay}$, 
when the cosmic strings dissapear. 
The GW peak energy density and frequency at that time is roughly estimated as Eq.~(\ref{omega_gw3}) 
and $k_{\rm peak} \sim a H (\tau_{\rm decay}) \simeq a(t_{\rm decay}) c_H^{-1/2} \mphi$, respectively. 
Then the GW amplitude decreases with time as $\propto a^{-1} (\tau)$ 
until reheating completes.

\vspace{0.2cm}
\noindent {\bf Results of numerical simulations. }
To calculate the GW spectrum from Eqs.~(\ref{omega_gw}), (\ref{A}), and (\ref{B}), 
we have performed lattice simulations with $N^3 = 256^3$ grid points 
in the oscillation era, in which $H (\tau) \propto a^{-3/2}$. 
We use a numerical method similar to the one used in Ref.~\cite{Hiramatsu:2013qaa}. 
While details are given in Ref.~\cite{full paper}, 
we stress that our results are independent of the choices of the time step, simulation size, and grid size 
by changing their values by a factor of $50\%$. 
Initial fluctuations of the field value are seeded by the vacuum fluctuations 
though we have checked that our results are qualitatively insensitive to the detailed form of the initial conditions. 
Numerical simulations are performed in a unit with $a (\tau_i) \equiv a_i = 1$ and $H(\tau_i) \equiv H_i = 1$, 
though we explicitly write $H_i$ below.

Figure~\ref{fig1} shows evolution of GW spectra 
obtained from our numerical simulations. 
We confirm that cosmic string network reaches scaling regime 
at the time around $\tau H_i \simeq 15$. 
Then the produced energy density of GWs $\Delta \Omega_{\rm gw} / \Delta \log \tau$ becomes constant 
while its peak frequency $k_{\rm peak}$ decreases with time as $k_{\rm peak} \propto \tau^{-1}$. 
When the flat direction starts to oscillate at the time around $\tau H_i = 35$ 
($\simeq \tau_{\rm decay} H_i$), 
the peak frequency becomes constant and the GW spectrum begins to redshift adiabatically 
as $\propto \tau^{-2}$ ($\propto a^{-1}$). 
We find that GW spectra for $k \lesssim \tau_{\rm decay}$ are consistent with analytical estimation 
with $\TT = \text{const.}$ (red dashed curve), 
which we expect from the viewpoint of causality. 
From Fig.~\ref{fig1}, we obtain numerical factors such that 
\beq
  \Omega_{\rm gw} ( t_{\rm decay} )
 &\simeq&  
 2 \lmk \frac{\la \phi \ra}{\M} \rmk^4, 
 \label{omega_gw4} \\ 
 \frac{k_{\rm peak}}{a(t_{\rm decay})} 
 &\simeq& 3 \frac{\mphi}{\sqrt{c_H}}, 
 \label{f_0}
\eeq
where $\la \phi \ra$ and $t_{\rm decay}$ are given by Eqs.~(\ref{VEV}) and (\ref{t_decay}), respectively.

\vspace{0.2cm}
\noindent {\bf Present spectrum of GWs and detectability. }
Since the GW energy density decrease with time as $\propto a^{-1} (\tau)$ 
in the oscillation era, i.e., matter dominated era, 
its present value is given as 
\beq
 && \Omega_{\rm gw} h^2 (t_0) \nonumber\\ 
 &\simeq& 
 \Omega_r h^2 
 \lmk \frac{g_s (t_0)}{g_s(t_{\rm RH})} \rmk^{4/3} 
 \lmk \frac{g_*(t_{\rm RH})}{g_* (t_0)} \rmk 
 \lmk \frac{H_{\rm RH}}{H_{\rm decay}} \rmk^{2/3}
 \Omega_{\rm gw} ( t_{\rm decay}) \nonumber \\
 &\simeq& 2 \times 10^{-7}
 \lmk \frac{\mphi }{10^3 \GeV} \rmk^{-2/3} \lmk \frac{\TR}{10^9 \GeV} \rmk^{4/3} 
 \lmk \frac{M_*}{\M} \rmk^{10/3}, \nonumber \\ 
 \label{result omega}
\eeq
where $t_0$ is the present time, 
$\Omega_r h^2$ ($\simeq 4.15 \times 10^{-5}$) is the present energy density of radiation, 
and $g_*$ ($g_s$) is the effective relativistic degrees of freedom for the energy (entropy) density. 
We have used Eq.~(\ref{omega_gw4}) 
and assumed $c'_H = a'_H = 1$ and $n=4$ in the last line (see Eqs.~(\ref{c_H}), (\ref{a_H}), and (\ref{VEV})). 
Taking redshift into account, 
we obtain the present value of peak frequency $f_0$ like 
\beq
 f_0 
 &\simeq& 
 \lmk \frac{g_s (t_0)}{g_s(t_{\rm RH})} \rmk^{1/3} 
 \lmk \frac{T_0}{T_{\rm RH}} \rmk 
 \lmk \frac{H_{\rm RH}}{H_{\rm decay}} \rmk^{2/3}
 \frac{k_{\rm peak}}{2 \pi a(t_{\rm decay})} \nonumber \\
 &\simeq& 7 \times 10^2 \text{ Hz} \lmk \frac{\mphi}{10^3 \GeV} \rmk^{1/3} \lmk \frac{\TR}{10^9 \GeV} \rmk^{1/3} 
  \lmk \frac{M_*}{\M} \rmk^{1/3}, \nonumber \\
 \label{result f}
\eeq
where we use Eq.~(\ref{f_0}) in the last line. 
As explained above, 
the spectrum of GWs bends at the wavenumber corresponding to the Hubble scale 
at the time of reheating ($k = k_{\rm bend} \simeq a(t_{\rm RH}) H_{\rm RH}$). 
The present value of bending frequency $f_{\rm bend}$ is given as 
\beq
 f_{\rm bend} 
 &=& 
 \lmk \frac{g_s (t_0)}{g_s(t_{\rm RH})} \rmk^{1/3} 
 \lmk \frac{T_0}{T_{\rm RH}} \rmk 
 \frac{k_{\rm bend}}{2 \pi a(t_{\rm RH})} \nonumber\\ 
 &\simeq& 30 \text{ Hz} 
 \lmk \frac{\TR}{10^9 \GeV} \rmk. 
 \label{f_bend}
\eeq
We can obtain the reheating temperature $\TR$ 
from observation of the bend in the GW spectrum at the large scale 
as observation of the inflationary GW background~\cite{Nakayama:2008ip}. 
In addition, we can obtain the mass of the flat direction $\mphi$ and the cut-off scale $M_*$ 
from observations of energy density and peak frequency of GWs (see Eqs.~(\ref{result omega}) and (\ref{result f})).

\begin{figure}[t]
\centering 
\includegraphics[width=.45\textwidth, bb=0 0 360 339]{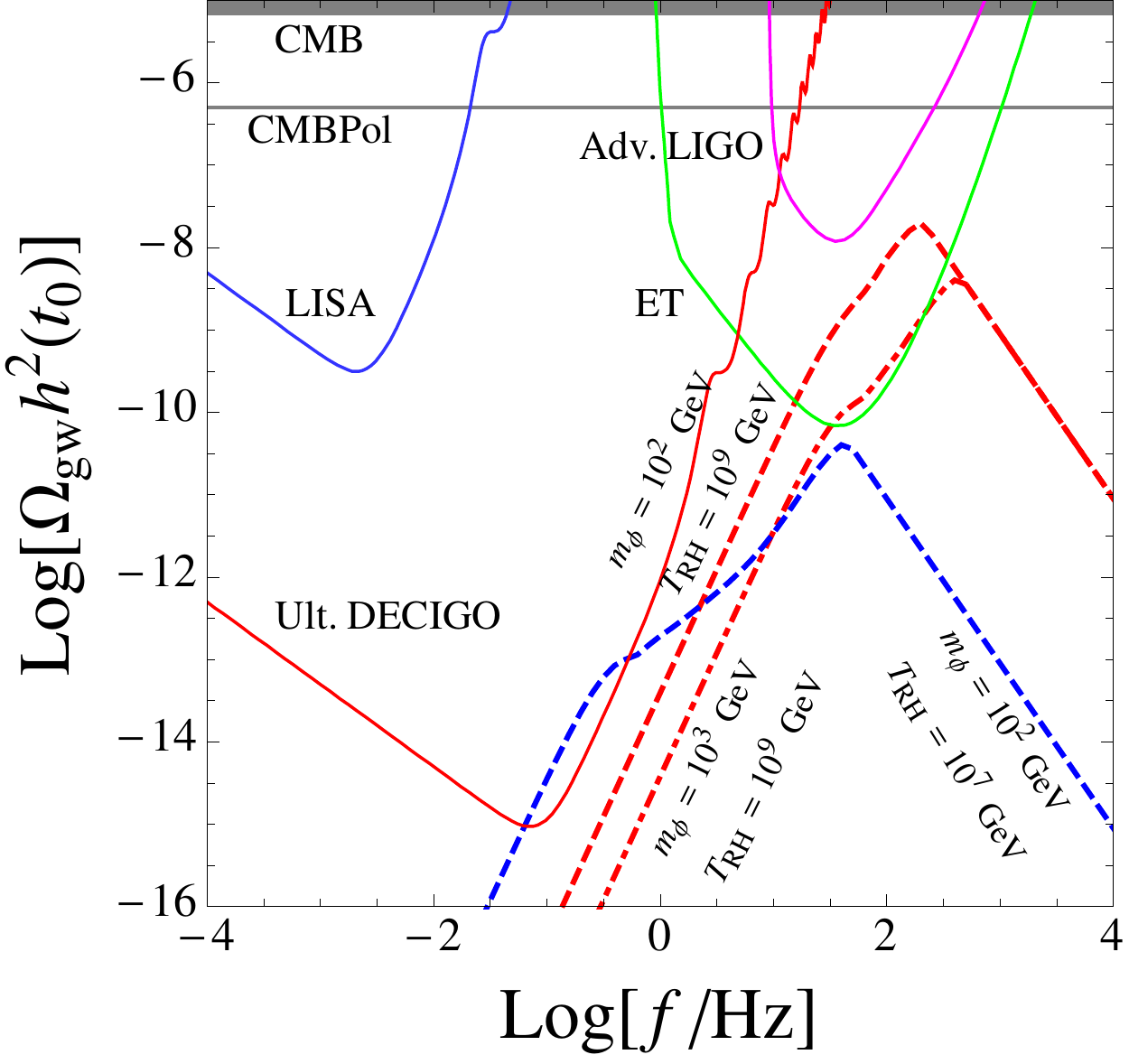} 
\caption{
GW spectra generated by cosmic strings 
and sensitivities of planned interferometric detectors. 
We plot the cases with 
$\mphi = 10^2 \GeV$ (red dashed curve) 
and $\mphi = 10^3 \GeV$ (red dot-dashed curve) 
for $\TR = 10^9 \GeV$. 
We also plot the case with $\mphi = 10^2$ and $\TR = 10^7 \GeV$ (blue dashed curve). 
We have assumed $M_*^2/\M^2 = 0.1$, $n=4$, and $c'_H = a'_H = 1$. 
}
  \label{fig2}
\end{figure}

Figure~\ref{fig2} shows examples of GW spectra 
predicted by the present mechanism. 
We also plot single detector sensitivities for LISA~\cite{lisa} 
and Ultimate DECIGO~\cite{decigo} 
by using the online sensitivity curve generator in~\cite{sens_curves} 
with the parameters in Table $7$ of Ref.~\cite{Alabidi:2012ex}. 
We plot cross-correlation sensitivities for Advanced LIGO~\cite{adv_ligo} 
and ET (ET-B configuration)~\cite{et}, 
assuming two detectors are co-alighned and coincident. 
In the figure, we take the signal to noise ratio ${\rm SNR} = 5$, 
the angular efficiency factor $F=2/5$, 
the total observation time $T=1 \,{\rm yr}$, and the frequency resolution $\Delta f / f =0.1$. 
CMB constraints (horizontal lines) are put on the integrated energy density of GWs 
$\int \dd \log f \Omega_{\rm gw} h^2 (t_0)$~\cite{Smith:2006nka}. 
We find that GW signals can be observed by ET and Ultimate DECIGO.

%
\section*{Acknowledgement}
M.Y. thanks Keisuke Harigaya, Masahiro Ibe, and Masahiro Kawasaki for useful comments. 
This work is supported by
the World Premier International Research Center Initiative (WPI Initiative), MEXT, Japan (A.K. and M.Y.),
the Program for Leading Graduate Schools, MEXT, Japan (M.Y.),
and JSPS Research Fellowships for Young Scientists (M.Y.).
%



\end{document}